\documentclass[12pt,preprint]{aastex}

\newcommand{\etal}{{et al.~}}

\newcommand{\lta}{\lesssim}
\newcommand{\gta}{\gtrsim}

\newcommand{\kms}{\>{\rm km}\,{\rm s}^{-1}}
\newcommand{\Gyr}{\>{\rm Gyr}}

\newcommand{\Mpc}{\>{\rm Mpc}}
\newcommand{\Myr}{\>{\rm Myr}}
\newcommand{\Msun}{\>{\rm M_{\odot}}}

\slugcomment{ApJ Letters, in press}

\shorttitle{RGB detection in SBS~1415+437}
\shortauthors{Aloisi et al.}

\begin{document}


\title{Do young galaxies exist in the Local Universe? ---\\ Red Giant
Branch detection in the metal-poor dwarf SBS~1415+437$^1$}


\author{A.~Aloisi\altaffilmark{2,3}, R.~P.~van der Marel\altaffilmark{2}, 
J.~Mack\altaffilmark{2}, C.~Leitherer\altaffilmark{2}, 
M.~Sirianni\altaffilmark{2,3} and M.~Tosi\altaffilmark{4}}


\altaffiltext{1}{Based on observations with the NASA/ESA Hubble
Space Telescope, obtained at the Space Telescope Science Institute,
which is operated by AURA for NASA under contract NAS5-26555.}

\altaffiltext{2}{Space Telescope Science Institute, 3700 San Martin Drive, 
Baltimore, MD 21218; aloisi@stsci.edu, marel@stsci.edu, mack@stsci.edu, 
leitherer@stsci.edu, sirianni@stsci.edu}

\altaffiltext{3}{On assignment from the Space Telescope Division of the 
European Space Agency.}

\altaffiltext{4}{INAF-Osservatorio Astronomico di Bologna, Via Ranzani 1, 
I-40127 Bologna, Italy; monica.tosi@bo.astro.it}


\begin{abstract}
We present results from an HST/ACS imaging study of the metal-poor
blue compact dwarf galaxy SBS~1415+437. It has been argued previously
that this is a very young galaxy that started to form stars only $\lta
100 \Myr$ ago. However, we find that the optical color-magnitude
diagram prominently reveals asymptotic giant branch and red giant
branch (RGB) stars. The brightness of the RGB tip yields a distance $D
\approx 13.6 \Mpc$. The color of the RGB implies that its stars must
be older than $\sim 1.3 \Gyr$, with the exact age depending on the
assumed metallicity and dust extinction. The number of RGB stars
implies that most of the stellar mass resides in this evolved
population. In view of these and other HST results for metal-poor
galaxies it seems that the local Universe simply may not contain any
galaxies that are currently undergoing their first burst of star
formation.
\end{abstract}





\section{Introduction}

Within the framework of hierarchical formation, dwarf ($M$ $\lesssim$
10$^9$ M$_{\odot}$) galaxies are often considered the first systems to
collapse, supplying the building blocks for the formation of more
massive galaxies through merging and accretion. So present-day dwarfs
may have been sites of the earliest star-formation (SF) activity in
the Universe. However, this hypothesis is challenged by the physical
properties of blue compact dwarf (BCD) galaxies, which have blue
colors indicative of a young stellar population. BCDs are experiencing
intense SF (0.01-10 M$_{\odot}$ yr$^{-1}$; Thuan 1991) but still have
a high neutral gas content ($\gtrsim$ 10$^8$ M$_{\odot}$; Thuan \&
Martin 1981). Oxygen abundances in H~{\sc ii} regions generally imply
a low metal content, with values in the range $12+\log({\rm O/H}) =
7.1$--$8.3$ (Izotov \& Thuan 1999; hereafter IT99). Some BCDs contain
much less heavy elements than the majority of high-$z$ galaxies. The
most metal-poor BCDs ($12+\log({\rm O/H}) \lta 7.6$) have therefore
been pointed out as particularly good candidate ``primeval'' galaxies
in the nearby Universe that started to form stars no more than $40
\Myr$ ago (IT99). This would support the view that SF in low-mass
systems has been inhibited until the present epoch (Babul \& Rees
1992). In any case, chemically poorly evolved star-forming dwarfs are
the best laboratory where to study SF processes similar to those
occurring in the early Universe, but with a spatial resolution and
sensitivity that are impossible to achieve in high-$z$ galaxies.

The most direct way to infer the age of a nearby galaxy is to resolve
it into individual stars and study the color-magnitude diagram (CMD).
The red giant branch (RGB) sequence, formed by evolved stars with ages
in excess of $\sim 1 \Gyr$, is of particular interest. An unambiguous
RGB detection implies that SF was already active more than a Gyr ago,
whereas absence of the RGB indicates that the system has started
forming stars only recently. Here we report results from a new $V$ and
$I$-band imaging study of the metal-poor BCD SBS~1415+437, performed
with the Hubble Space Telescope (HST) Advanced Camera for Surveys
(ACS).

Together with I~Zw~18 (see Section~\ref{s:conc}), SBS~1415+437 may be
one of the best candidate primeval galaxies in the local Universe. It
has an elongated shape with a bright H~{\sc ii} region at its SW tip
(see Figure~\ref{f:ACSimage}). The systemic velocity corrected for
Local Group infall towards Virgo is $851 \kms$ (from the Lyon
Extragalactic Database). Ignoring any possible peculiar
velocity, this implies a distance $D = 12.2 \Mpc$ and an intrinsic
distance modulus $m-M = 30.42$ for $H_0 = 70 \kms \Mpc^{-1}$. The
galaxy has an oxygen abundance $12+\log({\rm O/H}) = 7.59 \pm 0.01$
(IT99) and was studied in considerable detail by Thuan, Izotov, \&
Foltz (1999, hereafter T99). It has a blue integrated color, $V-I \lta
0.4$ throughout the main body, indicating that young stars are
dominating its light. Its SF regions were studied through UV and
optical spectroscopy, stellar population synthesis modeling and
chemical abundance determination. T99 concluded from all the combined
evidence that SBS~1415+437 is a young galaxy that did not start to
make stars until $\sim 100 \Myr$ ago. T99 also presented HST/WFPC2
data that resolved the brightest young stars, but these data were too
shallow (1800 sec in $V$ and 4400 sec in $I$) to reach evolved
asymptotic giant branch (AGB) or RGB stars.

\section{Observations and Data Reduction}

SBS~1415+437 was observed in HST GO program 9361 using the ACS Wide
Field Channel (WFC). The full field of view is $\sim 200 \times 200$
arcsec$^2$ with a pixel size of $0.05 \times 0.05$ arcsec$^2$. Total
integration times of $20160$ sec were obtained for each of the filters
F606W and F814W. The 16 optimally dithered single exposures per filter
were reprocessed with the most up-to-date version of the ACS
calibration pipeline. The exposures were registered, corrected for
geometric distortion, and co-added with cosmic-ray rejection using the
{\tt MultiDrizzle} software. Figure~\ref{f:ACSimage} shows a true
color image created by combining the data from the two filters.

Point source photometry was performed with the {\tt DAOPHOT/ALLSTAR}
package (Stetson 1987). A spatially variable point-spread function
(PSF) was inferred from the most isolated stars in the target vicinity
and was fitted to the detected stars within the galaxy. A master
catalog was created by matching peaks in excess of $4\sigma$ in a sum
of the $V$ and $I$ images, yielding $\sim 21000$ detected
objects. Because of the large number of dithered exposures, the master
catalog is essentially free of instrumental artifacts such as cosmic
rays or hot pixels. The vast majority of the detected sources are
individual stars in the galaxy, with a small potential contamination
from stars blends and background galaxies (foreground star
contamination is predicted to be negligible). We did not remove
potentially contaminating sources from our catalog by applying cuts in
sharpness or $\chi^2$, so as to not run the risk of rejecting bona
fide stars in SBS~1415+437 as well. However, we did experiment with
rejection strategies of various kinds and found that none of the
results presented here depend on this aspect of the analysis.

Photometry was performed at the positions of the objects in the master
catalog separately in the $V$ and $I$ band images. Aperture
corrections were measured from stars in the frame, and applied to all
photometry. Corrections for imperfect Charge Transfer Efficiency (CTE)
were applied based on extrapolation of the calibration of Riess \&
Mack (2004). Contamination to F606W by extended H$\alpha$ emission was
corrected in the photometric analysis through local background
subtraction. Count rates were transformed to the Johnson-Cousins $V$
and $I$ magnitudes using the color-dependent synthetic transformations
given by Sirianni et al.~(2005), which have been shown to be accurate
to $\sim 0.02$ mag.

Figure~\ref{f:CMDs}a shows the inferred $(V-I,\> I)$ CMD for the full
area of the sky covered by the galaxy. The median random errors in the
photometry are indicated in the figure. These errors do not account
for the possible influence of star blending, which will generally add
additional systematic uncertainties. An estimate of the 50\%
completeness limit is also shown in the figure. This estimate was
obtained by fitting the combination of a parameterized completeness
function and a model SF history to the observed CMD. The resulting
estimate is an average over the galaxy, given that the actual
completeness varies spatially depending on the level of crowding. The
simple estimates of the errors and completeness in
Figure~\ref{f:CMDs}a suffice for the purposes of the present paper. In
a forthcoming follow-up paper we will present artificial star tests to
estimate these quantities more robustly, and also as function of
position in the galaxy. That paper will also contain more details
about the data reduction and photometric analysis, and will discuss
narrow-band imaging obtained in the context of the same HST program.

\section{CMD Analysis}

At bright magnitudes in the CMD ($I \lta 25$) we find the same star
types detected previously with WFPC2 (T99): main sequence stars, and
their more evolved counterparts, the young supergiants. The combined
feature of main sequence stars and blue supergiants (evolved stars at
the hot edge of their core He-burning phase) shows up near $V-I
\approx 0.0$, while red supergiants appear near $V-I \approx 1.3$. Our
deeper ACS data show, as expected, that these sequences extend down to
much fainter magnitudes. But more importantly, the ACS data reveal the
unmistakable presence of a much older population. AGB and carbon stars
are seen at $I=24$--$26$, with colors extending from $V-I \approx 1.2$
to as much as $V-I \approx 4.0$. Below $I \approx 26.5$ there is a
prominent RGB with a pronounced RGB tip (TRGB).

We used the software developed by one of us (RvdM) and described in
Cioni \etal (2000) to determine the TRGB magnitude: $I_{\rm TRGB} =
26.67 \pm 0.02 ({\rm random}) \pm 0.03 ({\rm systematic})$. We also
determined the color of the TRGB, using a histogram of the star colors
in a magnitude range just below the TRGB: $(V-I)_{\rm TRGB} = 1.28 \pm
0.02 ({\rm random}) \pm 0.03 ({\rm systematic})$. The systematic
errors include the uncertainty introduced by the transformations from
the ACS filters to the Johnson-Cousins bands, the algorithmic
uncertainties associated with finding a discontinuity in the
luminosity function, and the fact that this discontinuity is not
infinitely steep due to observational uncertainties.

To fit the RGB we use theoretical isochrones from the Padua group, as
compiled and transformed to the Johnson-Cousins system by Girardi
\etal (2002).\footnote{The models and their properties are listed in
rows 2--8 of Table~1 in Girardi \etal (2002). They are available
electronically at http://pleiadi.pd.astro.it/ . The models all have
solar abundance ratios; $\alpha$-enhanced models are more appropriate
for BCDs, but are not generally available at the metallicities
relevant for BCDs.} The models include a simple synthetic scheme for
TP-AGB evolution. Their metallicity $Z$ (the mass fraction of metals
heavier than helium) samples the range from $0.0001$ to $0.03$. To
constrain the age of the RGB stars with the help of these models we
need some independent knowledge of the metallicity and the internal
reddening $A_V$.

If SBS~1415+437 has the same ratio of oxygen to metals as the Sun,
then the observed nebular oxygen abundance can be combined with the
most up-to-date solar values $12+\log({\rm O/H})_{\odot} = 8.66$ and
$Z_{\odot} = 0.0122$ (Asplund, Grevesse \& Sauval 2005) to obtain $Z =
0.0010$. However, if the galaxy has the same ratio of iron to metals
as the Sun then the implied metallicity is only $Z = 0.0003$, because
the HII regions in SBS~1415+437, like other BCDs, are
$\alpha$-enhanced compared to the Sun (T99).  Either way, the RGB
stars probably formed from gas that was more pristine and less
metal-enriched than the present-day gas for which nebular abundances
are available. We therefore adopt $Z = 0.0010$ as a fairly firm upper
limit to the metallicity of SBS~1415+437.

T99 used the Balmer decrement in a nuclear spectrum to estimate an
internal reddening $A_V = 0.25$, corresponding to $E(V-I) = 0.13$. We
adopt this as an upper limit to the average extinction suffered by RGB
stars, for three different reasons: (i) the nebular gas is usually
associated with SF regions, which tend to be inherently more dusty
than the regions in which older stars reside (as demonstrated
explicitly for the case of the LMC; Zaritsky 1999); (ii) the nuclear
regions of galaxies often tend to be the most dusty ones; and (iii)
the T99 estimate did not account for possible Balmer absorption
contamination from an underlying evolved population. Given the upper
limit to $E(V-I)$, and taking into account the random and systematic
errors, we infer that the true unreddened color of the TRGB must be in
the range $1.10 \leq (V-I)_{\rm TRGB} \leq 1.33$.

Figure~\ref{f:TRGBcolor} shows contours of the predicted TRGB color in
the parameter space of metallicity vs.~age. The TRGB becomes redder if
either age or metallicity is increased. The filled green area shows
the region bounded by the available constraints on $Z$ and $(V-I)_{\rm
TRGB}$. It shows that the age of the observed red giants might be
anywhere between $\sim 1.3 \Gyr$ and the Hubble time. To obtain the
youngest ages, one must assume that the dust extinction $A_V$ and
metallictity $Z$ are at the upper extremes of their allowed ranges.

For reference it is useful to consider one particular model in more
detail. Let us choose the model with $Z = 0.0010$ and $A_V = 0$ and
call this the ``standard'' model. To reproduce the observed color of
the TRGB in this model, the age of the RGB stars must be $2.2
\Gyr$. The absolute magnitude of the TRGB at this age is $M_{I,{\rm
TRGB}} = -3.99$. This implies a distance modulus $m-M = 30.66$ ($D
\approx 13.6 \Mpc$).  Figure~\ref{f:CMDs}a overplots in red the RGB
for the $2.2 \Gyr$ isochrone on top of the data, showing an excellent
fit. Figure~\ref{f:CMDs}b plots isochrones for other ages in the
standard model and shows good qualitative agreement also with the
other sequences in the observed CMD. Only the carbon star sequence is
not well fit, which is a known shortcoming of these stellar evolution
models (Marigo, Girardi \& Chiosi 2003).

The inferred TRGB distance modulus for the standard model has a
systematic error of at least $0.1$ mag due to uncertainties in the
evolutionary calculations (e.g., Bellazzini \etal 2004). There is an
additional systematic uncertainty associated with the unknown dust
extinction $0 \leq A_I \lta 0.12$. So overall the result agrees well
with the value $m-M = 30.42$ implied by straightforward application of
the Hubble law. It also agrees with a simple estimate that can be
obtained from the carbon stars. Their modal I-band magnitude is $I =
25.78 \pm 0.04$ (see Fig.~\ref{f:CMDs}). If we assume that these
carbon stars are on average equally luminous as those in the LMC at
$m-M = 18.5 \pm 0.1$ (van der Marel \& Cioni 2001), then SBS~1415+437
must have $m-M = 30.45 \pm 0.11 ({\rm random})$. The age and
metallicity dependence of the carbon star magnitudes is poorly known,
but probably adds at least $\sim 0.2$ mag of systematic error to this
estimate. Either way, these agreements provide an independent reason
why the RGB stars in SBS~1415+437 cannot be younger than $\sim 1.3
\Gyr$. Not only would such young RGB stars have a TRGB that is bluer
than observed, but $M_{I,{\rm TRGB}}$ would also be up to 2 mag
fainter than the usual value (Barker, Sarajedini \& Harris 2004) since
in the youngest RGB stars He ignites in non-degenerate cores. The
implied TRGB distance modulus of SBS~1415+437 would then be quite
inconsistent with the values implied by the Hubble law and the
observed carbon star brightnesses. Moreover, for ages below $\sim 1.3
\Gyr$ the RGB and its tip cease to be well-defined features in the
CMD, in conflict with the observed CMD morphology in
Figure~\ref{f:CMDs}a.

At $I=27$ the observed RGB has a Gaussian $V-I$ dispersion of $0.19$
(see Fig.~\ref{f:CMDs}a), which is twice the median photometric
error. This might be because the photometric random errors
underestimate the true errors. Alternatively, the observed RGB width
may indicate an intrinsic spread in properties. For the standard
model, the observed RGB width can be reproduced either with a Gaussian
dispersion of $0.60$ dex in $Z$ or $0.31$ dex in age. A spread in the
extinctions $E(V-I)$ towards individual stars might also contribute
towards the observed width.

The mass in evolved stars can be estimated using the best-fitting
($2.2 \Gyr$) isochrone in the standard model. For a Salpeter Initial
Mass Function (IMF) from $0.1$--$100 \Msun$, only 1 in $2.4 \times
10^4$ randomly drawn stars is observed in a 1 magnitude range below
the TRGB. Since we actually observe $\sim 2500$ stars there (with $1.0
\leq V-I \leq 1.5$) the galaxy must have $\sim 6 \times 10^{7}$ stars
on this isochrone. The associated mass is $2.1 \times 10^7 \Msun$ (the
average mass per star is $0.35 \Msun$ for the adopted IMF). The mass
increases if the RGB stars are assumed to be older. T99 estimated the
mass in young stars ($\lta 100 \Myr$) to be only $1.2 \times 10^6
\Msun$. Detailed modeling of the SF history will be presented in a
forthcoming paper, but preliminary results broadly confirm that at
least 80\% of the stellar mass of SBS~1415+437 resides in stars with
ages $\gta 1.3 \Gyr$. The follow-up paper will also address the
variations in the SF history along the main body of the galaxy, which
was previously found by T99 to be quite non-homogeneous. Here we
simply note that the RGB stars seen in the CMD of Figure~\ref{f:CMDs}
are in fact spread over the entire galaxy, and are not just found
exclusively in one particular region.

\section{Conclusions \& Discussion}
\label{s:conc}

We have used HST/ACS to detect AGB and RGB stars in the metal-poor BCD
SBS~1415+437. The data imply that most of the stellar mass of this
galaxy resides in stars older than $\sim 1.3 \Gyr$. It was proposed
previously that this galaxy (T99), and others like it (IT99), did not
form stars more than $\sim 100 \Myr$ ago. This was based primarily on
interpretation of integrated spectra and heavy element abundances in
HII regions. Our results show that such data should be used with
caution when addressing the SF history of the underlying stellar
population.

Our results add to the growing list of low-metallicity galaxies in
which an RGB has been detected with HST, including the BCDs I~Zw~36
($12+\log({\rm O/H}) = 7.77$; $D = 5.9 \Mpc$; Schulte-Ladbeck \etal
2001), VII Zw 403 ($12+\log({\rm O/H}) = 7.69$; $D = 4.5 \Mpc$;
Schulte-Ladbeck \etal 1999) and UGC~4483 ($12+\log({\rm O/H}) = 7.54$;
$D = 3.2 \Mpc$; Dolphin et al.~2001; Izotov \& Thuan 2002), and the
Local Group dwarf irregular galaxies Leo A ($12+\log({\rm O/H}) =
7.30$; Tolstoy et al.~1998, Schulte-Ladbeck et al.~2002) and SagDIG
($12+\log({\rm O/H}) = 7.26$--$7.50$; Momany \etal 2005). The
situation for the most metal-poor BCD, I~Zw~18, has been less
clear-cut, possibly because it is more metal-poor than any other
galaxy studied ($12+\log({\rm O/H}) = 7.18 \pm 0.01$; IT99) or because
of its larger distance ($D \approx 15 \Mpc$). Early HST images did
reveal asymptotic giant branch (AGB) stars in I~Zw~18 (e.g., Aloisi,
Tosi \& Greggio 1999; \"Ostlin 2000), but Izotov \& Thuan (2004) did
not detect an RGB in more recent deeper imaging with ACS. However, the
latter result has now been challenged by Momany \etal (2005) based on
a better photometric analysis of the same data. They show that many
red sources do exist at the expected position of an RGB, and that
their density in the CMD drops exactly where a TRGB would be
expected. Additional HST/ACS data of I Zw 18 may be needed for a
conclusive understanding of its SF history. But either way, the
preponderance of the evidence now seems to suggest that the local
Universe simply may not contain any galaxies that are currently
undergoing their first burst of star formation.



\acknowledgments

Support for proposal 9361 was provided by NASA through a grant from
STScI, which is operated by AURA, Inc., under NASA contract NAS
5-26555. E.~Smith and T.~Brown are acknowledged for their advice on the
photometric reduction of crowded fields observed with ACS.

\clearpage


\begin{figure}
\begin{center}
{\tt This figure is available as a gif file from astro-ph}
\end{center}
\caption{True-color composite image of SBS~1415+437 showing a field of
view of $52.5 \times 33.5$ arcsec$^2$ centered on the target. HST/ACS
F606W data ($V$) are shown in blue and F814W data ($I$) in red. North
is at about $-14.6^{\circ}$ from horizontal. The inset in the bottom
right shows a single-band blow-up of a $1 \times 1$ arcsec$^2$ region
to give a sense of the amount of crowding and noise. This region is
neither a best-case nor a worst-case in terms of crowding, but is
fairly typical for the main body of the galaxy.
\label{f:ACSimage}}
\end{figure}

\begin{figure*}
\begin{center}
{\tt This figure is available as a gif file from astro-ph}
\end{center}
\caption{(a) CMD of the resolved stellar population of
SBS~1415+437. The data are corrected for foreground extinction, but
not for possible extinction internal to the galaxy. The median
photometric random error as a function of $I$-band magnitude is
indicated on the right. The dashed curve provides an estimate of the
50\% completeness level. The RGB for an age of $2.2 \Gyr$ is
overplotted in red for our ``standard model'' with metallicity $Z =
0.0010$, no internal extinction, and distance modulus $m-M
=30.66$. (b) Padua model isochrones for the standard model (black) for
(roughly from top left to bottom right) $\log ({\rm age}) = 6.75$,
$7.00$, $7.25$, $\ldots$, $9.75$, $10.00$. The main evolutionary
sequences seen in the data are indicated in approximate sense as
colored straight lines: main sequence (MS), blue supergiants (BSG),
red supergiants (RSG), the red giant branch (RGB) with its tip (TRGB),
the asymptotic giant branch (AGB) and carbon stars (AGB stars in which
carbon has been dredged up to the surface).
\label{f:CMDs}}
\end{figure*}

\begin{figure}
\plotone{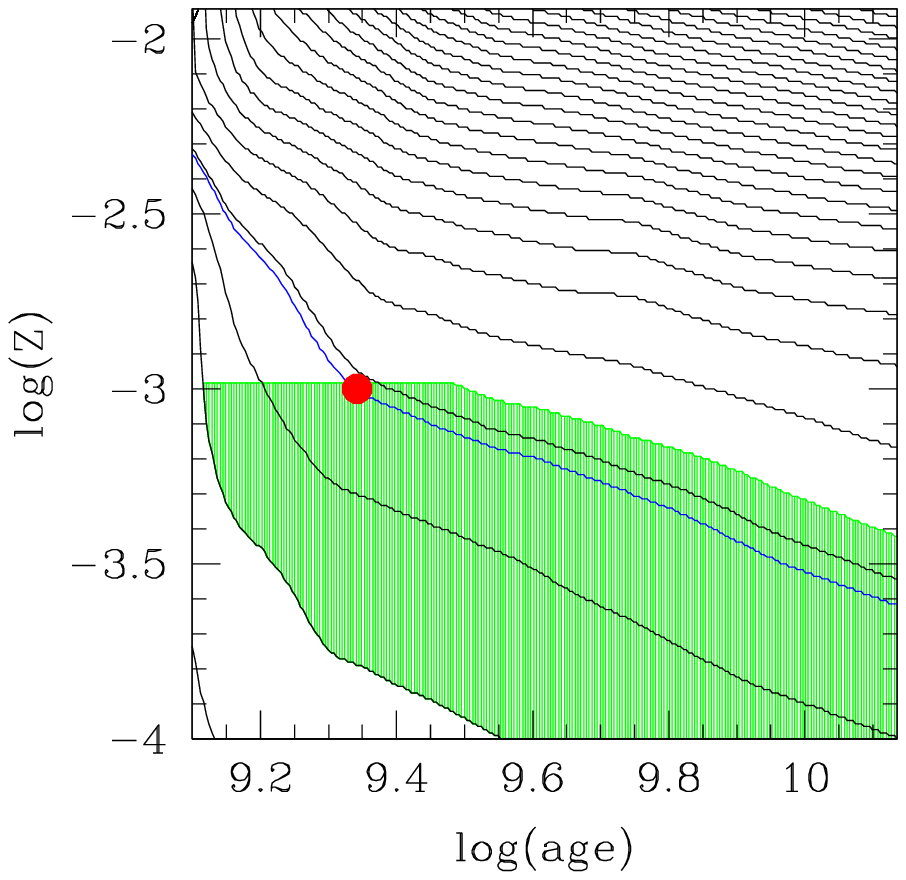}
\caption{Curved solid lines show contours of constant TRGB color in
the parameter space of $\log(Z)$ vs.~$\log({\rm age})$. The age is
expressed in years. The metallicity $Z$ runs from $0.0001$ at the
bottom of the plot to $0.0122$ (the solar metallicty $Z_{\odot}$;
Asplund \etal 2005) at the top of the plot. The contours were
calculated from the Padua model isochrones and run from $(V-I)_{\rm
TRGB} = 1.0$ in the bottom left to $(V-I)_{\rm TRGB} = 3.6$ in the top
right, in steps of $0.1$. The blue curve corresponds to the actually
observed color $(V-I)_{\rm TRGB} = 1.28$. The red dot on this curve
marks the parameter combination for which the RGB is shown in
Figure~\ref{f:CMDs}a. The green area marks the region in which the RGB
stars of SBS~1415+437 must reside, given the available constraints on
its metallicity and dust extinction.
\label{f:TRGBcolor}}
\end{figure}


\begin{thebibliography}{}

\bibitem[]{Alo99} Aloisi, A., Tosi, M., \& Greggio, L.~1999, AJ, 118, 302

\bibitem[]{Asp05} Asplund, M., Grevesse, N., \& Sauval, J. 2005, in
"Cosmic abundances as records of stellar evolution and nucleosynthesis", F.N. Bash \& T.G Barnes, eds., ASP, in press [astro-ph/0410214]

\bibitem[]{Bel04} Bellazzini, M., Ferraro, F.~R., Sollima, A., Pancino, E., \&
Origlia, L. 2004, A\&A, 424, 199

\bibitem[]{Bab92} Babul, A., \& Rees, M.~J.~1992, MNRAS, 255, 346

\bibitem[]{Bar04} Barker, M. K., Sarajedini, A., \& Harris, J. 2004, ApJ, 
606, 869

\bibitem[]{Cio00}
Cioni, M. R., van der Marel, R. P., Loup, C., \& Habing, H. J. 2000, 
A\&A, 359, 601

\bibitem[]{Dol01} Dolphin, A.~E., et al.~2001, MNRAS, 324, 249

\bibitem[]{Gir02} Girardi, L., Bertelli, G., Bressan, A., Chiosi, C., 
Groenewegen, M.A.T., Marigo, P., Salasnich, B., \& Weiss, A. 2002, A\&A 391

\bibitem[]{Izo99} Izotov, Y.~I., \& Thuan, T.~X.~1999, ApJ, 511, 639 (IT99)

\bibitem[]{Izo02} Izotov, Y.~I., \& Thuan, T.~X.~2002, ApJ, 567, 875

\bibitem[]{Izo04} Izotov, Y.~I., \& Thuan, T.~X.~2004, ApJ, 616, 768 

\bibitem[]{Mar03} Marigo P., Girardi L., Chiosi C., 2003, A\&A, 403, 225

\bibitem[]{Mom05} Momany, Y., et al. 2005, A\&A, in press [astro-ph/0505399]

\bibitem[]{Ost00} \"Ostlin, G.~2000, ApJ, 535, L99

\bibitem[]{Rie04} Riess, A., \& Mack, J.~2004, ACS Instrument Science Report 
2004-006 (Baltimore: STScI)

\bibitem[]{Sch99} Schulte-Ladbeck, R.~E., Hopp, U., Crone, M.~M., 
\& Greggio, L.~1999, ApJ, 525, 709

\bibitem[]{Sch01} Schulte-Ladbeck, R.~E., Hopp, U., 
Greggio, L., Crone, M.~M., Drozdovsky, I.~O.~2001, AJ, 121, 3007

\bibitem[]{Sch02} Schulte-Ladbeck, R.~E., Hopp, U., Drozdovsky, I.~O., 
Greggio, L., \& Crone, M.~M.~2002, AJ, 124, 896

\bibitem[]{Sir05} Sirianni, M., et al.~2005, PASP, submitted

\bibitem[]{Ste87} Stetson, P.~B.~1987, PASP, 99, 191

\bibitem[]{Thu91} Thuan, T.~X.~1991, in Massive Stars in Starbursts, 
     ed.~C.~Leitherer, N.~R.~Walborn, T.~M.~Heckman, \& C.~A.~Norman 
     (Cambridge Univ.~Press), 183

\bibitem[]{Thu99} Thuan, T.~X., Izotov, Y.~I., \& Foltz, C.~B.~1999, ApJ, 
     525, 105 (T99)

\bibitem[]{Thu81} Thuan, T.~X., \& Martin, G.~E.~1981, ApJ, 247, 823

\bibitem[]{Tol98} Tolstoy, E., et al.~1998, AJ, 116, 1244

\bibitem[]{vdM01} van der Marel, R. P., \& Cioni, M.-R. L. 2001, AJ, 122, 1807 

\bibitem[]{Zar99} Zaritsky, D. 1999, AJ, 118, 2824

\end{thebibliography}
\end{document}